\documentclass[prl,aps,showpacs,groupedaddress,superscriptaddress,twocolumn,toc=flat,longbibliography]{revtex4-1}

\usepackage[colorlinks=true,linkcolor=blue,urlcolor=blue,citecolor=blue]{hyperref}

\usepackage[utf8x]{inputenc}
\usepackage{color}
\usepackage[usenames,dvipsnames]{xcolor}
\usepackage{bbm} 

\usepackage{amsfonts,ams math,amssymb,stmaryrd}

\usepackage{graphicx}
\usepackage{subfigure}  
\usepackage{bbm} 
\usepackage{hyperref}
\usepackage{epsfig}
\usepackage{mathrsfs}
\usepackage{verbatim}
\usepackage{centernot}
\usepackage{ulem}
\usepackage{mathtools}
\usepackage{nicefrac}

\renewcommand{\l}{\left(}
\renewcommand{\r}{\right)}

\renewcommand{\H}{\hat{\mathcal{H}}}

\renewcommand{\c}{\hat{c}}

\newcommand{\cd}{\hat{c}^\dagger}

\usepackage{array}

\usepackage{cancel,ifthen}
\newcommand{\cmnt}[2][NoInPuT]{\ifthenelse{\equal{#1}{NoInPuT}}{}{{\color{red}\sout{#1}}} {\color{blue} #2}}

\usepackage{bm}	
\renewcommand{\vec}[1]{\bm{#1}}

\begin{document}
\normalem	

\title{Classifying Snapshots of the Doped Hubbard Model with Machine Learning}

\author{Annabelle Bohrdt}
\affiliation{Department of Physics and Institute for Advanced Study, Technical University of Munich, 85748 Garching, Germany}
\affiliation{Department of Physics, Harvard University, Cambridge, Massachusetts 02138, USA}
\author{Christie S. Chiu}
\affiliation{Department of Physics, Harvard University, Cambridge, Massachusetts 02138, USA}
\author{Geoffrey Ji}
\affiliation{Department of Physics, Harvard University, Cambridge, Massachusetts 02138, USA}
\author{Muqing Xu}
\affiliation{Department of Physics, Harvard University, Cambridge, Massachusetts 02138, USA}
\author{\\ Daniel Greif}
\affiliation{Department of Physics, Harvard University, Cambridge, Massachusetts 02138, USA}
\author{Markus Greiner}
\affiliation{Department of Physics, Harvard University, Cambridge, Massachusetts 02138, USA}
\author{Eugene Demler}
\affiliation{Department of Physics, Harvard University, Cambridge, Massachusetts 02138, USA}
\author{Fabian Grusdt}
\affiliation{Department of Physics and Institute for Advanced Study, Technical University of Munich, 85748 Garching, Germany}
\affiliation{Department of Physics, Harvard University, Cambridge, Massachusetts 02138, USA}
\author{Michael Knap}
\affiliation{Department of Physics and Institute for Advanced Study, Technical University of Munich, 85748 Garching, Germany}

\pacs{}

\date{\today}

\begin{abstract}

Quantum gas microscopes for ultracold atoms can provide high-resolution real-space snapshots of complex many-body systems. We implement machine learning to analyze and classify such snapshots of ultracold atoms. Specifically, we compare the data from an experimental realization of the two-dimensional Fermi-Hubbard model to two theoretical approaches: a doped quantum spin liquid state of resonating valence bond type, and the geometric string theory, describing a state with hidden spin order. This approach considers all available information without a potential bias towards one particular theory by the choice of an observable and can therefore select the theory which is more predictive in general. Up to intermediate doping values, our algorithm tends to classify experimental snapshots as geometric-string-like, as compared to the doped spin liquid. Our results demonstrate the potential for machine learning in processing the wealth of data obtained through quantum gas microscopy for new physical insights.
\end{abstract}

\maketitle

\paragraph{Introduction.--}\label{secIntro}The phase diagram of the Fermi Hubbard model and its connection to high-temperature superconductivity have been the subject of a vast amount of theoretical and experimental studies in the past decades~\cite{Lee2006,Keimer2015}. While a large number of theories exist, each with its own merits, a unifying analytic understanding is nonetheless still lacking. In the regime of low temperatures and finite doping even numerical simulations become increasingly difficult. 
In recent years, tremendous progress has been made in using ultracold atoms to study quantum magnetism in the Fermi-Hubbard model~\cite{Greif2013, Hart2015,Cheuk2016,Hilker2017,Mazurenko2017,Brown2018,Nichols2018,Salomon2018,Chiu2018}. These ultracold atom experiments are now exploring  finite doping regimes of the phase diagram where no consensus on a theoretical description and the most appropriate way to experimentally characterize the system exists.
\\
All information about the quantum state of the system is contained in the many-body density matrix, where the number of degrees of freedom scales exponentially with the system size. A measurement collapses the quantum state, such that only a projection of it can be accessed. Repeated projective measurements provide a plethora of data, which in the past has mostly been analyzed to obtain conventional observables such as one- and two-point correlation functions, that are also traditionally measured in solids. However, measurements performed in quantum gas microscopes contain considerably more information. 
Therefore, the need arises for new methods to analyze the data that take all available information into consideration and hence use the capabilities of quantum gas microscopes to their full extent. 
\onecolumngrid

\vskip .6cm
\begin{figure*}[h]
	\centering
	\epsfig{file=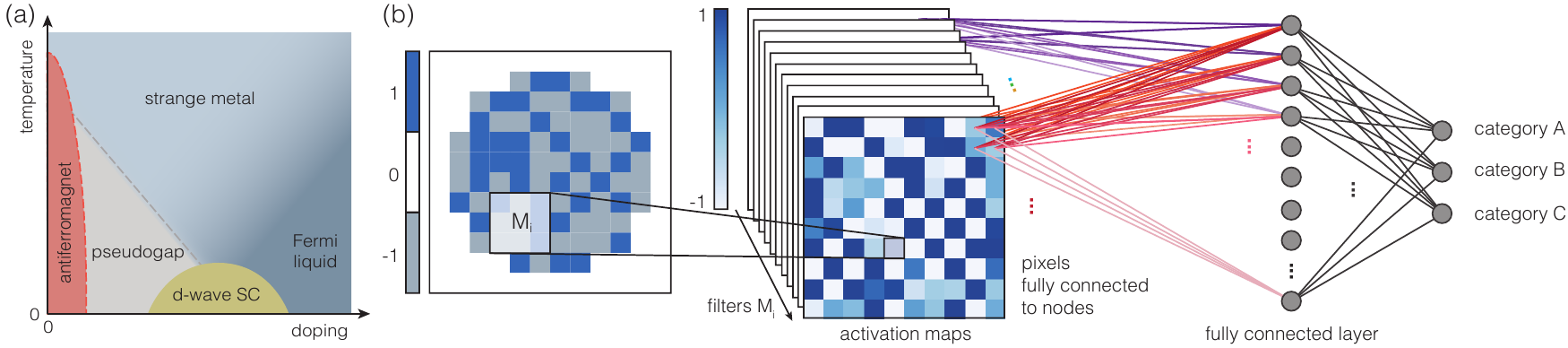}
	\caption{\textbf{Classifying quantum gas microscope snapshots of the doped Fermi-Hubbard model with convolutional neural networks (CNNs).} (a) Schematic phase diagram of the 2D Fermi-Hubbard model. We use snapshots of the many-body quantum state at fixed doping and temperature as input data for the CNN. (b) The main building block of CNNs, which are conventionally used to analyze visual imagery, is the convolutional layer with a set of learnable filters $\mathrm{M_i}$ as parameters~\cite{Goodfellow2016}.  At each possible position of a given filter in the input image, the inner product between the filter and the input data is computed. This yields a two dimensional activation map of the filter. During training, the network learns to set the entries of the filters such that the corresponding value in the activation map is high when specific types of patterns are detected. The convolutional layer is followed by a fully connected layer, which then sorts the data into the different categories.
	}
	\label{fig1Schematics}
\end{figure*}

\twocolumngrid
In this work, we employ machine learning techniques to characterize the large amount of data produced by quantum gas microscopy of the doped Fermi-Hubbard model. Recently, machine learning has emerged as a new tool in condensed matter physics. Its main applications so far include representing the wave functions of correlated many-body states \cite{Carleo2017,Glasser2017,Lu2018}, the determination and characterization of a phase transition~\cite{Carrasquilla2017, Rem2018, Broecker2017, Chng2017, Beach2018, Dong2018,Greitemann2018,Liu2018,Koch-Janusz2017} and most recently the analysis of scanning tunneling microscopy images~\cite{Zhang2018}. 
Here, we take a complementary route and use a CNN to classify experimental data at finite doping into different theoretical categories in order to determine which theory describes the system best on the microscopic level, see Fig.~\ref{fig1Schematics}. This approach provides insights into the underlying microscopic structures of the state, which may be inaccessible to conventional observables but can be essential for gaining a deeper understanding of the emergent physics. 
\paragraph{Physical system.--}The experimental data that we analyze with our machine-learning algorithm has been measured with a quantum gas microscope for ultracold Lithium atoms in an optical lattice and is available in Ref.~\onlinecite{Chiu2018}. This system is modeled by the Fermi-Hubbard Hamiltonian
\begin{equation}
\H = -t \sum_{\sigma = \uparrow, \downarrow} \sum_{\langle \vec{i}, \vec{j} \rangle} \l \cd_{\vec{i},\sigma} \c_{\vec{j},\sigma} + {\rm h.c.} \r+ U \sum_{\vec{j}} \cd_{\vec{j},\uparrow} \c_{\vec{j},\uparrow} \cd_{\vec{j},\downarrow} \c_{\vec{j},\downarrow},
\label{eqFH}
\end{equation}
where the first term describes tunneling with amplitude $t$ of spin-$1/2$ fermions between nearest-neighbor sites of a two-dimensional square lattice. 
The second term corresponds to on-site interactions of strength $U$ between fermions with opposite spin; $U\approx 8t$ in the experiment~\cite{Chiu2018}. 
The half-filling limit of the two-dimensional Hubbard model is comparably well understood and can be approximately described for large interactions by the Heisenberg Hamiltonian with superexchange coupling $J = 4t^2/U$~\cite{Auerbach1998}. Starting from high temperatures $T>J$, upon decreasing the temperature, anti-ferromagnetic (AFM) correlations with increasing correlation length emerge.
We now investigate the decrease of AFM correlations with doping by comparing the snapshots obtained from the quantum gas microscope to two different theories, a doped resonating valence bond (RVB) liquid \cite{Anderson1987,Baskaran1987} and the geometric string theory \cite{Grusdt2018tJz,Grusdt2018mixD,Chiu2018} over a wide range of dopings. 
Before presenting our results, we provide a brief account of the two theories from which we numerically sample snapshots of the many-body density matrix. 
\begin{figure}[t!]
\centering
\epsfig{file=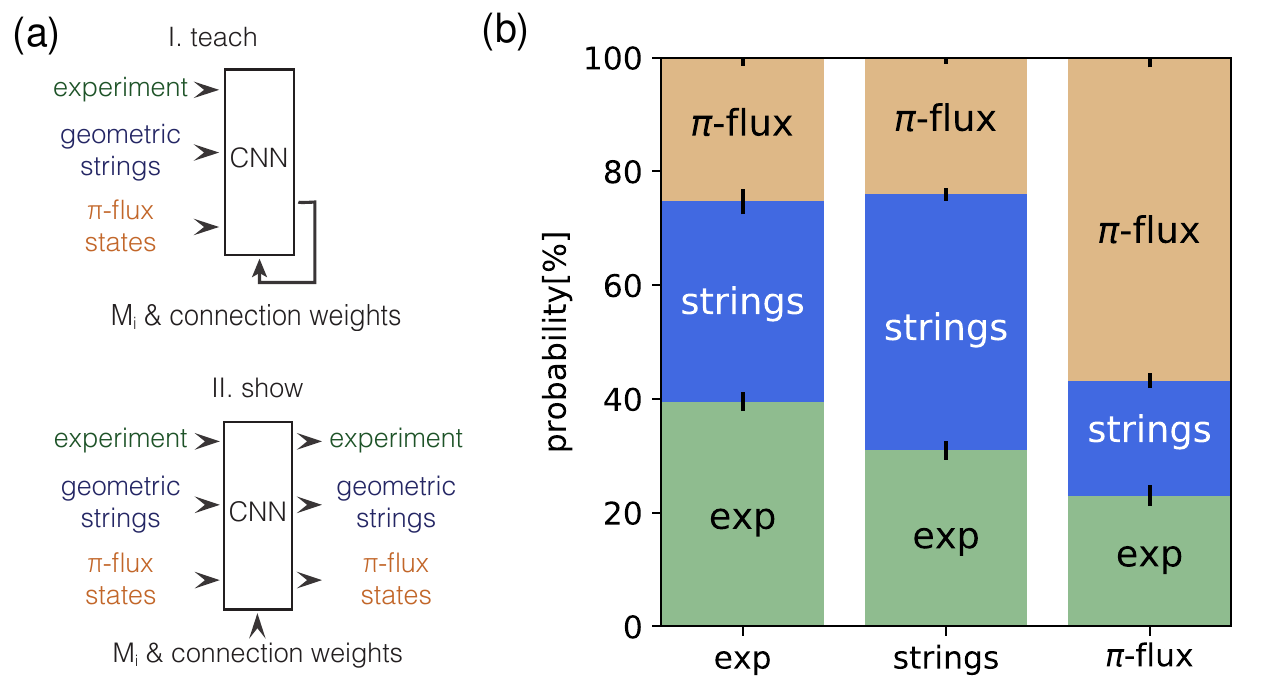, width=\columnwidth}
\caption{\textbf{Classifying single snapshots of the many-body density matrix.} (a) The convolutional neural network is trained to identify to which dataset category any given snapshot belongs. Here, we consider (i) experimental data, (ii) geometric strings and (iii) snapshots from $\pi$-flux theory, all at 9\% doping. The probabilities in (b) show how a sample of 400 snapshots which have not been used during training is classified. While the $\pi$-flux theory is recognized comparably well, a clear separation between experimental data and geometric strings is not possible. We repeat the process ten times to average out fluctuations in the results, see Supplementary Material for details~\cite{supp}.
}
\label{fig1All}
\end{figure}
\paragraph{$\pi$-flux theory.--}
\label{secPFT}
In the RVB picture, the ground state of the doped Hubbard model is described as a superposition of different spin-singlet coverings of the lattice, through which deconfined chargons can move freely. Our simulations for this theory are based on a mean-field parton Hamiltonian $\H_{\rm MF}$ with free spin-$\nicefrac{1}{2}$ fermions hopping on a square lattice with a magnetic flux of $\pi$ per plaquette \cite{Anderson1987,Baskaran1987,Marston1989}. A Gutzwiller projection of the corresponding thermal density matrix $\hat{\rho} = e^{- \beta \H_{\rm MF}}$ removes double occupancies in accordance with a large on-site interaction $U\gg t$. 
\\
We use Monte Carlo sampling techniques to generate snapshots in the Fock basis of the projected mean-field density matrix. To take into account virtual charge fluctuations present in the larger physical Hilbert space, we introduce doublon-hole pairs into the snapshots on neighboring sites with probability $4t^2/U^2$ determined by second order perturbation theory. 
The overall energy scale in the mean-field Hamiltonian is fixed such that the nearest-neighbor spin correlator at half-filling matches the experimental value. This approach has been shown in Ref. \onlinecite{Chiu2018} to lead to good agreement of spin correlations for all relevant doping values. Our results are robust under small variations in the overall energy scale.
\paragraph{Geometric string theory.--}
\label{secGST}
In the underdoped regime, this theory describes the fermionic charge carriers as bound states of two partons \cite{Beran1996,Baskaran2007,Punk2015PNASS}: a heavy spinon and a light chargon; see also Refs.~\cite{Bulaevskii1968,Trugman1988,Manousakis2007}. Their internal structure is described by a fluctuating geometric string of displaced spins connecting the spinon to the chargon \cite{Grusdt2018tJz,Grusdt2018mixD}. 
In order to derive the properties of this string, the frozen spin approximation is assumed, in which the spin background does not change with doping but the anti-ferromagnetic order is hidden by the hole motion. 
\\
Each hole displaces the spins along the string by one site, which leads to an increase in spin interaction energy proportional to the spin correlations in the undoped system and a decrease of the overall staggered magnetization. The distribution of the geometric string length is obtained from a microscopic calculation of the motion of a single hole at a given temperature and Hubbard parameter $U/t$ \cite{Chiu2018}. 
\\
To generate snapshots for the geometric string theory, we start from the experimental data at half-filling and for each doping value place the corresponding number of holes independently into the snapshots. The holes are then moved indepenently from one another in random directions through the anti-ferromagnet for a number of sites which is sampled from the theoretical string length distribution.
\\
The experimental images only contain information about one spin species, while the other spin species as well as doublons and holes are detected as empty sites. Hence, before comparing our theoretical images to experimental results, the second spin species and doubly occupied sites are converted to empty sites in the theoretical data.
\paragraph{Classifying snapshots.--}
\label{secClassify1}

We now train a convolutional neural network to distinguish snapshots from the following classes: (i) experimental data, (ii) geometric string theory and (iii) $\pi$-flux theory, all at 9\% doping. In all snapshots, the underlying SU(2) symmetry of the Fermi-Hubbard Hamiltonian leads to fluctuations of the N\'eel ordering vector. To simplify the pattern recognition, we use $\sim 30\%$ of the snapshots with the highest absolute values of the staggered magnetization. 
Importantly, this still corresponds to a large number of snapshots, which can be considered as a representative sample of the quantum state, see Supplementary Material~\cite{supp}.
\\
The performance of our neural network is visualized in Fig.~\ref{fig1All}. In this plot, the $x$-axis displays the actual class of a snapshot and the $y$-axis shows the probability for the neural network to sort it into the different classes. The accuracy for the classification of images, which corresponds to the weighted average of the diagonal entries, is 47.1\%. The main source of confusion for the CNN is the similarity between the experimental and the geometric string theory data, while a differentiation of the $\pi$-flux theory snapshots is more successful. Taking the first two categories together, the accuracy of the classification increases to 69.2\%. This is a first indication that the geometric string theory resembles the experimental data at 9\% doping more closely than $\pi$-flux theory. 

\paragraph{Sorting experimental data into theory categories.--}
\label{secClassify2}
One of the most powerful features of neural networks is their ability to generalize to new situations not encountered during training. We make use of this property by first training a CNN to distinguish between snapshots from $\pi$-flux and geometric string theory at a fixed doping value; a task for which the CNN achieves a precision above 70\%. Subsequently we show experimental data to the CNN to sort it into one of the two theory categories. The classification of experimental data then reveals how similar these snapshots are to the theoretically simulated data. 
\begin{figure}[t!]
\centering
\epsfig{file=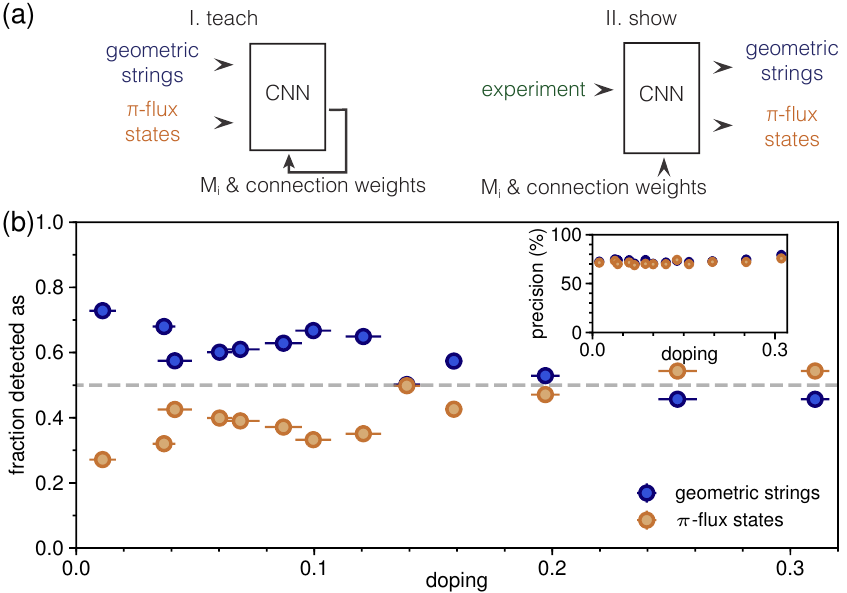, width=\columnwidth}
\caption{\textbf{Classifying experimental data.} (a) The network is trained to distinguish snapshots from geometric string theory (blue) and $\pi$-flux theory (orange) at each doping value. After training, experimental images at the same doping are shown to the network. (b) The average of the resulting classification of the experimental data at the corresponding doping value. 
The inset shows the precision for the trained classes on a subset of data not used for training. We repeat the process ten times to average out fluctuations in the results.
}
\label{fig2ASPi}
\end{figure}
\\
As shown in Fig.~\ref{fig2ASPi}, the neural network classifies a majority of the experimental snapshots as geometric string theory over a broad range of doping values up to about 15\%, even though conventional spin and charge correlation functions coincide equally well with experimental results in that regime for both theories~\cite{Chiu2018}. For larger dopings, the experimental data is more often sorted into the $\pi$-flux theory class, see also Supplementary Material~\cite{supp}.
\\
The ability of the neural network to distinguish $\pi$-flux from geometric string theory on the level of individual images indicates that the physical structure of these states is different. We can further improve the accuracy of our classification by taking into account the information that an entire set of measurements belongs to the same physical state. When the CNN sorts each snapshot into one of the two categories with probabilities $p$ and $1-p$, the entire sample is classified by the category in which the majority is sorted. Because the number of shots in each category follows a binomial distribution, the probability to make a wrong classification of the entire sample decays exponentially with the number of snapshots. Therefore, the entire experimental dataset at any doping value below $\sim15\%$ would be classified as geometric string theory data with almost 100\% probability~\cite{supp}. When the input to the network consists of four snapshots from the same category, the precision is above 80\% already and the fraction of experimental images classified as string theory increases significantly~\cite{supp}.

Moreover, our algorithm also classifies the low-temperature experimental snapshots to geometric string theory rather than experimental high temperature data, again up to doping values of about 15\%; see Supplementary Material for details.

\paragraph{Summary and Outlook.--}\label{secSummary}

We have demonstrated that convolutional neural networks provide a powerful tool to analyze the large amount of experimental data obtained from quantum gas microscopes. Individual snapshots can be classified to theoretical predictions and we can thereby determine which theory fits best. We apply this method to the Fermi-Hubbard model on a square lattice and find that on a microscopic level the experimental data more closely resembles the geometric string theory with short-range hidden order than the $\pi$-flux RVB theory in the regime of low doping.
Our analysis suggests a qualitative change of this behavior between 15\% to 20\% doping, beyond which the $\pi$-flux RVB data resembles the experiment more closely. 
\\
Conventional observables, such as the staggered magnetization or two-point spin correlation functions, hardly allow for a distinction between the theories under consideration and it depends on the chosen observable which theory will be favored~\cite{Chiu2018}. By contrast, the CNN searches for patterns in the collection of snapshots in an unbiased way without specifying certain physical observables and with that searches for structure in the many-body density matrix. Turning this argument around, it remains an interesting open challenge to understand how the CNN classifies the snapshots, which we plan to address in a future work.
\\
In this work, we compared two theories out of many potential candidates to the experimental data. In future work, the investigation of a larger class of theories will provide us with further information about the structure of the quantum state of the two-dimensional Fermi Hubbard model. Straightforward extensions include the comparison of snapshots from the Fermi-Hubbard model to different resonating valence bond states or predictions by quantum dimer models \cite{Punk2015PNASS}. Examining completely different parameter regimes or even models could reveal additional insights. Current experiments have been performed at comparably high temperatures, where no d-wave pairing or charge order is expected. Once colder temperatures are achievable, it will be interesting to compare geometric string theory to theoretical models with different types of order parameters built in. 
\\
The analysis of snapshots from quantum gas microscopy with machine learning techniques has the capability to reveal microscopic mechanisms and hidden order in the considerable amount of available data. Machine learning of quantum many-body states, perhaps possible through experimental snapshots, offers prospects to find the most predictive theory among a multitude of competing theories.
\paragraph{Acknowledgments.--}
We thank Ehud Altman, Juan Carrasquilla, Ehsan Khatami, Frank Pollmann, Achim Rosch, Subir Sachdev, Richard Schmidt and Dries Sels for insightful discussions and M\'arton Kan\'asz-Nagy additionally for his Heisenberg QMC code. We acknowledge support from Harvard-MIT CUA, NSF Grant No. DMR-1308435, AFOSR-MURI Quantum Phases of Matter (grant FA9550-14-1-0035), AFOSR grant No. FA9550-16-10323; DoD NDSEG; the Gordon and Betty Moore Foundation EPIQS program and grant No. 6791; NSF GRFP and grant Nos. PHY-1506203 and PHY-1734011; ONR grant No. N00014-18-1-2863; SNSF; Studienstiftung des deutschen Volkes; and the Technical University of Munich - Institute for Advanced Study, funded by the German Excellence Initiative and the European Union FP7 under grant agreement 291763, from the DFG grant No. KN 1254/1-1, and DFG TRR80 (Project F8) and the German Excellence Strategy MCQST.

\newpage
\clearpage
\appendix

\setcounter{figure}{0}
\setcounter{equation}{0}

\renewcommand{\thepage}{S\arabic{page}} 
\renewcommand{\thesection}{S\arabic{section}} 
\renewcommand{\thetable}{S\arabic{table}}  
\renewcommand{\thefigure}{S\arabic{figure}} 
\renewcommand{\theequation}{S\arabic{equation}}

\begin{center}
	\textbf{\Large{\large{Supplementary Material:\\Classifying Snapshots of the Doped Hubbard Model with Machine Learning}}}
\end{center}

\subsection{Data}

\begin{figure}[t!]
\centering
\epsfig{file=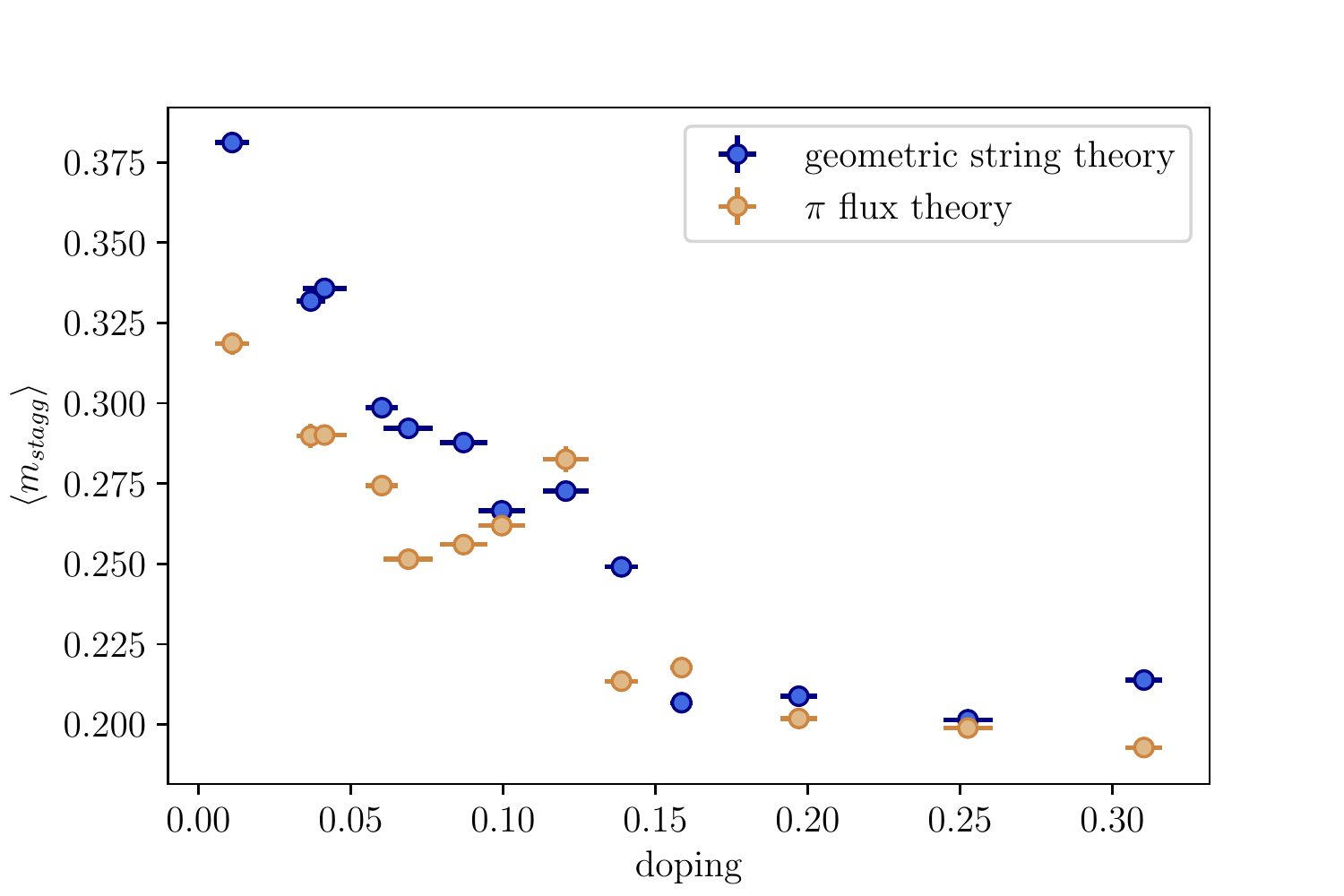, width=\columnwidth}
\caption{\textbf{Staggered magnetization of the two classes after sorting.} After training the CNN to distinguish geometric string theory and $\pi$-flux theory, we show it experimental data, which is then classified by the network as one of the two theories. As a result, we obtain a subset of experimental data sorted into the string theory class and accordingly a second complementary subset classified as $\pi$-flux theory. For these two new datasets, we now determine the average value of the staggered magnetization. The similarity of the resulting values for the different theories shows that a classification by the CNN solely based on the staggered magnetization is not convincingly possible.}
\label{figS2staggMag}
\end{figure}

The experimental data is obtained from Refs.~\onlinecite{Mazurenko2017,Chiu2018}. All snapshots used in this work consist of a circular region of interest (ROI) with 80 sites. In Table~\ref{table:snapshot_numbers} we list the amount of data available at $T=0.6J \pm 0.1J$ at various dopings, as well as at half-filling for different temperatures. 
\begin{table}[b!]
\caption{Number of snapshots from the experiment available at different doping values at $T=0.6J \pm 0.1J$ (left) and at different temperature values at half-filling (right). }
  \begin{tabular}{c| c || c | c }
  doping (\%) & number of snapshots & T/J & number of snapshots \\
    \hline
    \hline
        0 & 5326 & &  \\ \hline
    2 & 480 & 0.5 & 65 \\ \hline
        3 & 248 & 0.6 & 1776 \\ \hline
    4.5 & 253 & 0.7 & 4657 \\ \hline
    6 & 1588 & 0.8 & 356 \\ \hline
        7 & 687 & 0.9 & 346 \\ \hline
        9 & 1205 & 1.0 & 317 \\ \hline
    10 & 483 & 1.1 & 313 \\ \hline
    12.5 & 185 & 1.2 & 475 \\ \hline
        14 & 383 & 1.3 & 151 \\ \hline
        17 & 372 & 1.4 & 91 \\ \hline
    20 & 335 & 1.5 & 62 \\ \hline
    25 & 148 & 1.6 & 811 \\ \hline
        32 & 332 & 1.8 & 183 \\ 
  \end{tabular}
  \label{table:snapshot_numbers}
\end{table}
Conventional observables such as the staggered magnetization and spin correlators are very similar to the experiment for both the geometric string and the $\pi$-flux theories over a wide range of dopings~\cite{Chiu2018}. A distinction of the theories by the neural network solely based on such 'trivial' quantities can therefore be excluded. Furthermore, for the nearest-neighbor spin correlator, which could probably be used to differentiate single snapshots, the experiment seems to be better described by $\pi$-flux theory than by geometric string theory~\cite{Chiu2018}. However, geometric strings do describe the experimentally measured staggered magnetization more closely; as such we separately calculate the staggered magnetizations of the experimental data classified into geometric-string theory and into pi-flux theory in Fig.~\ref{figS2staggMag}. While the part of the dataset classified as $\pi$-flux theory has a slightly lower average staggered magnetization than the set of snapshots sorted into the geometric string theory class, see Fig.~\ref{figS2staggMag}, the values for both categories are comparably close to each other. The similarity of staggered magnetization values for both theories indicates that the CNN is classifying on the basis of more involved patterns. 
\\
As discussed in the main text, we post-select the snapshots in all cases to a high value of the staggered magnetization. For experimental as well as theory data, we calculate the absolute value of the staggered magnetization $|m_z|$ for every snapshot and use only the $\sim$30\% of all snapshots with the highest value of $|m_z|$. The post-selection is necessary to avoid confusion due to images with N\'eel ordering vector not aligned with the measurement axis, in which case a single snapshot alone does not contain any useful information. We performed the analysis shown in the main text for different percentages of the total amount of data and found that our results are independent of the details of the post-selection.
\subsection{Errors}
Before training the neural network, we randomly select 400 snapshots from the full data for testing. These snapshots are not used during training, but only to test the performance of the network and determine the accuracy. In order to average out fluctuations in the results, we repeat the training with a different choice of the test data set and sorting of experimental data 10 times. The error bars for the different accuracies as well as the percentage of experimental snapshots detect as one of the two theories shown in the figures represent the standard deviation over the different runs,
\begin{equation}
\Delta x = \frac{\sqrt{\sum_{i=1}^\mathcal{N} \left(x_i - \bar{x}\right)^2}}{\mathcal{N}-1},
\end{equation}
where $\mathcal{N}$ is the number of runs, $x_i$ is the quantity under consideration in the $i$-th run and $\bar{x}$ is the average value of the said quantity over all $\mathcal{N}$ runs. 

\subsection{Network}
The convolutional neural network used in this work was implemented using TensorFlow \cite{Tensorflow2015}. We tested the performance of the network with different architectures and while there are some quantitative fluctuations, the qualitative result is robust under changes of the architecture. Since the learning tasks as well as the amount of training data available are different for Fig.~\ref{fig1All}, Fig.~\ref{fig2ASPi} and Fig.~\ref{fig3AShot}, the optimal performance in terms of overall test accuracy is achieved for slightly different network architectures. The concrete parameters used for the different figures are listed in Table~\ref{table:network_parameters}. 
\begin{table}
\caption{Network parameters for Fig.~\ref{fig1All}, Fig.~\ref{fig2ASPi} and \ref{fig3AShot}. }
  \begin{tabular}{l || l | l | l  }
    & Fig.~\ref{fig1All} & Fig.~\ref{fig2ASPi} & Fig.~\ref{fig3AShot} \\
    \hline
    \hline
    available snapshots  & 2059 & 5326 & 2848 \\ \hline
    post-selection (\%) & 30 & 30 & 40 \\ \hline
    batch size  & 200 & 250 & 250 \\ \hline
   conv. layer 1  & 20  & 24 & 16  \\ \hline
   conv. layer 2   & - & - & 4 \\  \hline
   dropout   & 0.8 & 0.55 & 0.7  \\  \hline
   fully conn. layer 1  & 40 & 100 & 40 \\ \hline
   fully conn. layer 2  & 30 & - & - \\  \hline
   dropout   & 0.7 & 0.5 & 0.6  \\  \hline
   output layer & 3 & 2 & 2  \\ \hline
   learning rate   & 4 $\cdot 10^{-4}$ & 1 $\cdot 10^{-3}$ & 3 $\cdot 10^{-4}$ \\  \hline
    learning rate decay   & 0.6 & 0.5 & 0.5\\  \hline
    iterations  & 1000 & 1000 & 1000 \\
  \end{tabular}
  \label{table:network_parameters}
\end{table}

\subsection{Distinguishing theories}
\subsubsection{Geometric string theory based on Heisenberg QMC}
\begin{figure}[t!]
\centering
\epsfig{file=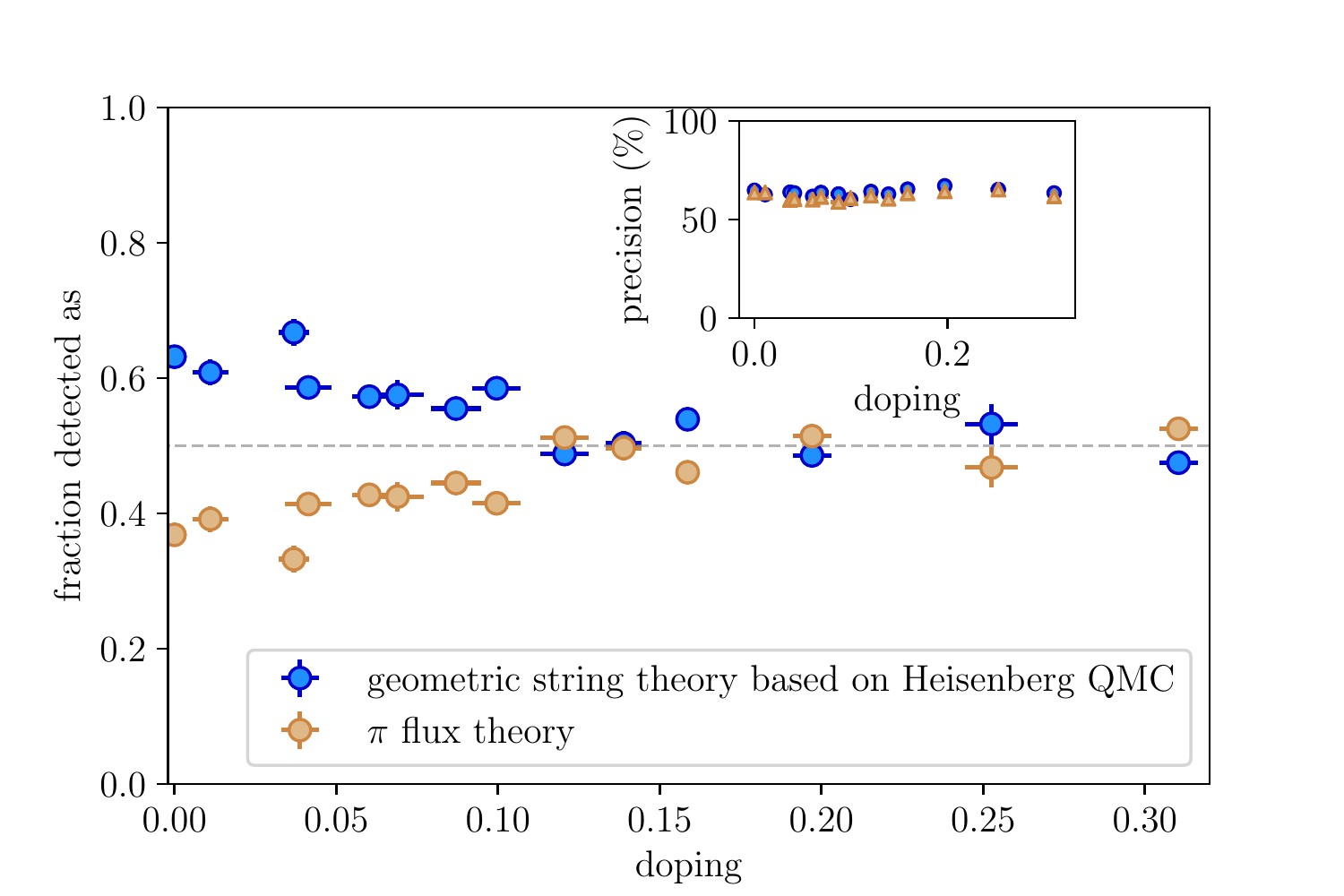, width=\columnwidth}
\caption{\textbf{Distinguishing geometric string theory from $\pi$-flux theory.} The network is trained to distinguish snapshots from geometric string theory (blue) and $\pi$-flux theory (orange) at each doping value. As opposed to Fig.~\ref{fig2ASPi} in the main text, the geometric string theory data is based on Heisenberg quantum Monte Carlo simulations to avoid a potential bias by using experimental half-filling data. After training, experimental images at the same doping are shown to the network. The plot shows the average of the resulting classification over all available experimental data at the corresponding doping value. The inset shows the precision that is achieved for the trained classes on a subset of data not used for training. We repeat the process ten times in order to average out fluctuations in the results. 
}
\label{figS2qmcPi}
\end{figure}
In the main text, we train the network to distinguish between geometric string theory and $\pi$-flux theory and subsequently sort experimental data into one of those categories. One potential bias of this approach is that the geometric string theory snapshots are based on experimental data at half-filling, whereas the $\pi$-flux theory data is purely theoretical apart from a fitted overall energy scale. This bias is to some extent intrinsic, because the geometric string theory does not make a statement about the state at half-filling itself, but only about how the introducing holes changes it. However, we can also generate purely theoretical images for the geometric string theory category by using Heisenberg Quantum Monte Carlo (QMC) snapshots as starting point. After adding the strings by hand as described above we also add doublon hole pairs in the same way as for $\pi$-flux theory. We then train the CNN to distinguish the geometric string theory data based on Heisenberg QMC from $\pi$-flux theory data and subsequently sort the experimental data into the two categories. As can be seen from Fig.~\ref{figS2qmcPi}, the qualitative result is very similar to what we obtained in Fig.~\ref{fig2ASPi} in the main text where geometric string theory data is based on experimental half-filling snapshots. 

\subsubsection{Distinguishing geometric string theory from random images}
\begin{figure}[t!]
\centering
\epsfig{file=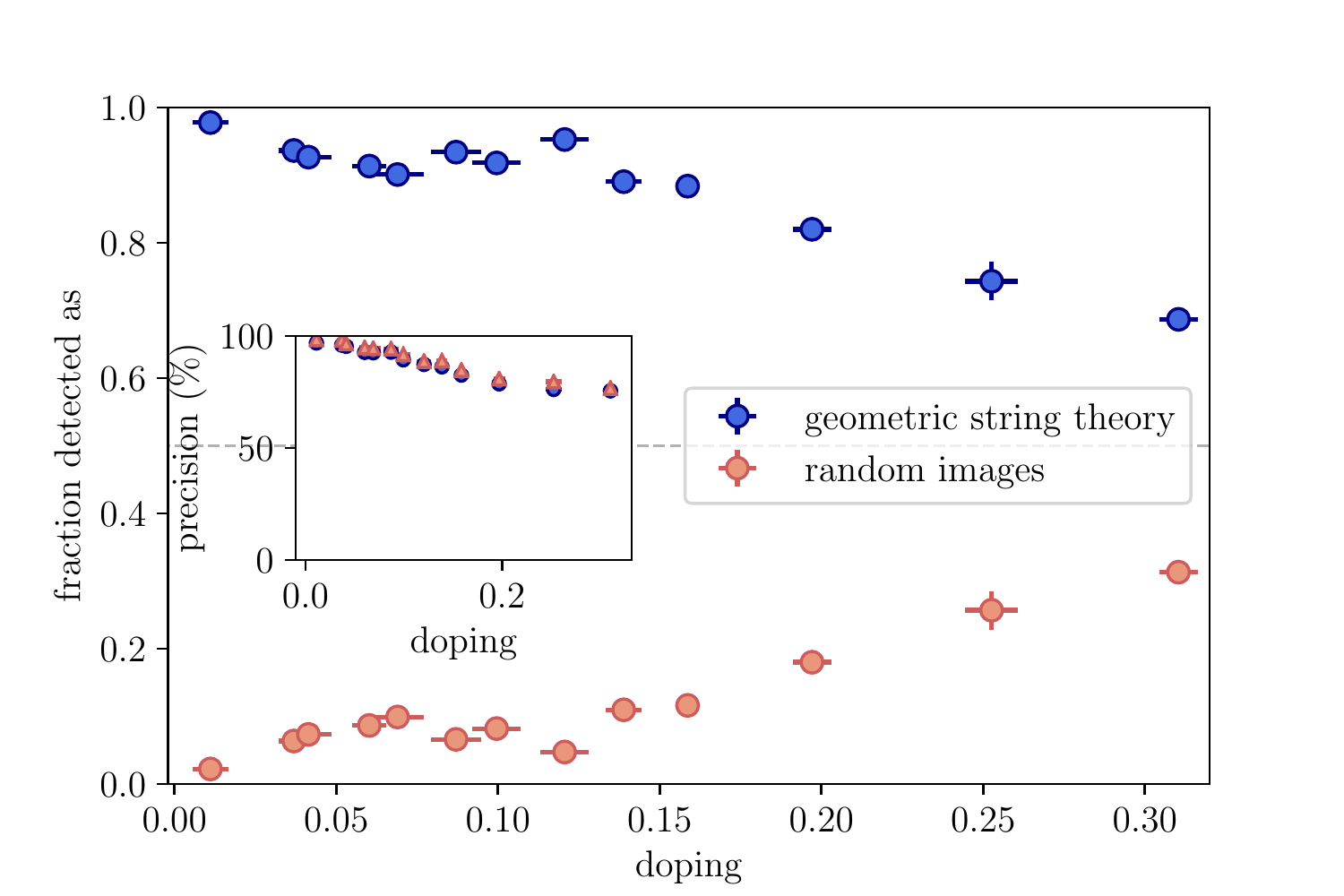, width=\columnwidth}
\caption{\textbf{Distinguishing geometric string theory from random snapshots.} The network is trained to distinguish snapshots from geometric string theory (blue) and random images (salmon) at each doping value. After training, experimental images at the same doping are shown to the network. The plot shows the average of the resulting classification over all available experimental data at the corresponding doping value. The inset shows the precision that is achieved for the trained classes on a subset of data not used for training. We repeat the process ten times to average out fluctuations in the results. 
}
\label{figS3ASrand}
\end{figure}
The precision achieved by the CNN during training for the different tasks so far has been around 75\%. In Fig.~\ref{figS3ASrand} we show the precision as well as the resulting classification of experimental data after training the network to distinguish geometric string theory from random data, where the ratio of occupied to empty sites in every image is chosen according to the doping value under consideration. For doping below 10\%, the precision is close to 100\%, demonstrating that the network is capable of learning to classify different theories almost perfectly.   
\subsubsection{Classifying sets of snapshots}
\begin{figure}[t!]
\centering
\epsfig{file=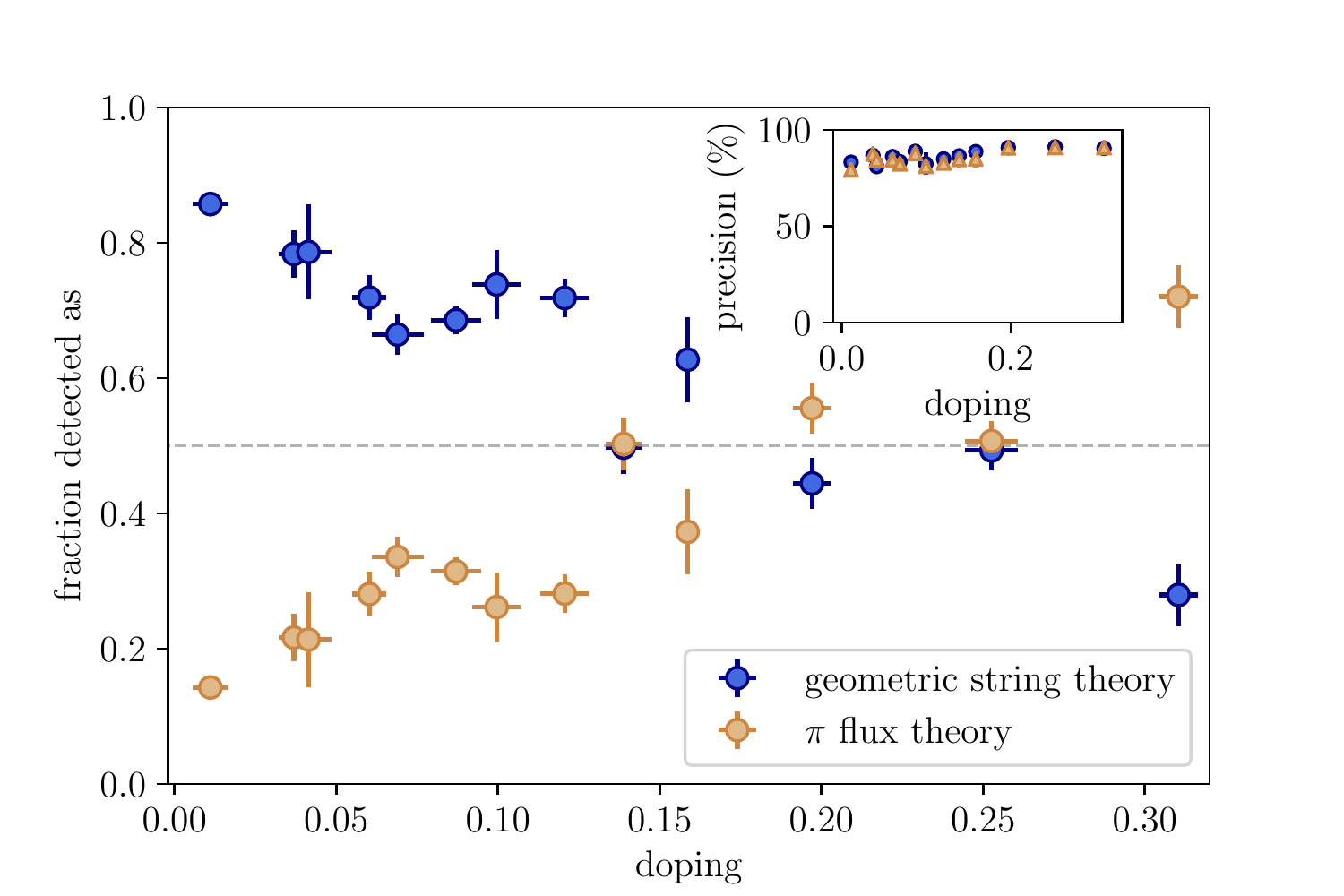, width=\columnwidth}
\caption{\textbf{Distinguishing geometric string theory from $\pi$-flux theory, four snapshots at a time.} The network is trained to distinguish snapshots from geometric string theory (blue) and $\pi$-flux theory (orange) at each doping value. Each input corresponds to four randomly chosen images from one theory. After training, experimental images at the same doping are shown to the network. Again, we show the network four images at a time and the network has to classify them into the same category. The plot shows the average of the resulting classification over all available experimental data at the corresponding doping value. The inset shows the precision that is achieved for the trained classes on a subset of data not used for training. We repeat the process five times to average out fluctuations in the results. 
}
\label{figS4ASpi4in1}
\end{figure}
In Fig.~\ref{fig2ASPi}, the neural network sorts about 60\% of the experimental snapshots into the geometric string theory category in the low doping regime. As stated in the main text, the accuracy of our classification can be increased further by taking into account the information that an entire set of measurements belongs to the same physical state. Here, we take one step in that direction and group four snapshots of the same category at a time. Additionally, for each of the four snapshots we choose a random number r = 0, 1, 2, or 3 and rotate the snapshot $r\cdot90$ degrees. Using the same network architecture, the precision increases to about 85\% throughout all doping values, see Fig.~\ref{figS4ASpi4in1}. After training, we show the network four experimental snapshots at a time and ask it to assign a single label to them. Again, we apply a random number of rotations to each individual snapshot. For dopings below 14\%, the fraction of experimental images classified as string theory is between 65 and 85\%.

\subsection{Comparing finite doping to high temperature}
\label{secClassify3} Starting at half-filling and cold temperatures around $0.6J$ the AFM correlations vanish both with increasing doping or increasing temperature~\cite{Chiu2018}. We approach the question of how similar these regimes are on the microscopic level. At each doping value we train the CNN to distinguish geometric string theory data and experimental data at high temperature and half-filling. We randomly add holes to the hot data according to the respective doping level to prevent the network from distinguishing the two theories trivially by the density. In order to obtain a sufficiently large training set, we use half-filling data for temperatures between $T=0.9J$ and $1.8J$ for the high temperature class.
\begin{figure}[t!]
	\centering
	\epsfig{file=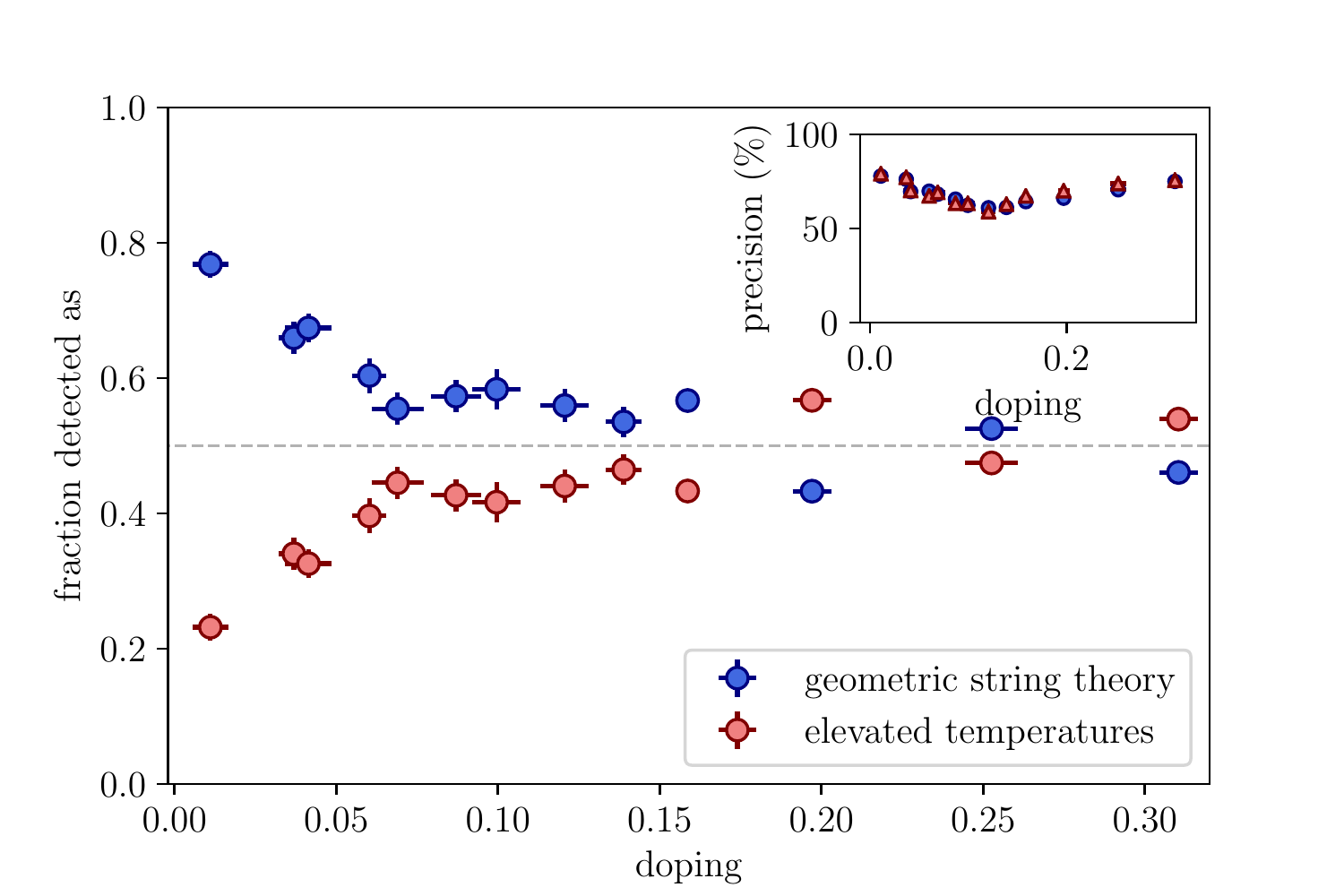, width=\columnwidth}
	\caption{\textbf{Distinguishing finite doping from high temperatures.} The CNN is trained to distinguish geometric string theory at a given doping value from experimental images at half-filling and temperatures $T\geq 0.9J$ with the corresponding number of holes randomly sprinkled into the pictures to prevent a distinction based on the filling. After training, experimental images at low temperatures, $T\approx 0.6J$, and the corresponding doping value are shown to the network. The plot shows the average of the resulting classification. The inset shows the precision that is achieved for the trained classes on a subset of data not used for training. The process is repeated ten times to average out fluctuations in the results. 
	}
	\label{fig3AShot}
\end{figure}
\\
For doping below $\sim15\%$, the network classifies the experimental data as geometric string theory. However, the precision decreases with increased doping, see inset of Fig.~\ref{fig3AShot}. This shows that it is difficult for the CNN to distinguish the theories, indicating that the classification of experimental pictures around 15\% doping is not very reliable. 

\newpage

\bibliography{CNN}{}

\end{document}